\documentclass[%
reprint,
superscriptaddress,
 amsmath,amssymb,
 aps,
pra,
l
]{revtex4-2}
\usepackage{xcolor}
\usepackage[normalem]{ulem}
\usepackage{graphicx}
\usepackage{dcolumn}
\usepackage{bm}
\usepackage{hyperref}
\hypersetup{
    colorlinks=true
}

\begin{document}

\preprint{APS/123-QED}

\title{Fr\"{o}hlich Condensate of Phonons in Optomechanical Systems}

\author{Xu Zheng}
\email{Xu.Zheng@Colorado.Edu}
\affiliation{Department of Physics, University of Colorado, Boulder, CO, 80309, USA}
\author{Baowen Li}
\email{Baowen.Li@Colorado.Edu}
\affiliation{Paul M. Rady Department of Mechanical Engineering, University of Colorado, Boulder, CO, 80309, USA}
\affiliation{Department of Physics, University of Colorado, Boulder, CO, 80309, USA}
\homepage{https://www.colorado.edu/faculty/li-baowen/}
\date{\today}

\begin{abstract}
 We propose that the Fr\"{o}hlich condensate of phonons can be realized in optomechanical systems. The system consists of a one-dimensional array of membranes coupled to the cavity field via a quadratic interaction, and the cavity is pumped by an external laser. Analytical and numerical results demonstrate that the high phonon occupancy of the lowest or highest mechanical mode is achievable depending on the detuning of the driving laser, the optomechanical strength, and the temperature. Feasibility of experimental implementation is discussed. Our results shed light on the energy conversion/transfer, heat control and multimode cooling using optomechanical systems.
\end{abstract}

\maketitle


\section{Introduction}
The study of open systems far from thermodynamic equilibrium has attracted attentions during the past decades. An interesting phenomenon in these systems is the emergence of collective behaviors and self-organization, which is the mechanism behind the generation of laser, superfluorescence \cite{bonifacio1975cooperative}, synchronization \cite{pikovsky2003synchronization,heinrich2011collective,zhang2015synchronization,colombano2019synchronization,sheng2020self}, etc. In biological systems, many theoretical works have suggested that the collective behavior may have profound effects on the chemical and enzyme kinetics \cite{reimers2009weak}, and the cognitive function of brain \cite{hameroff2014consciousness}. Among these works a widely used model is the Fr\"{o}hlich condensate \cite{frohlich1968long,frohlich1968bose,frohlich1970long}. In 1968, Fr\"{o}hlich showed that the energy of a collection of oscillators would concentrate at the lowest mode once the external energy supply exceeds a threshold. While this phenomenon is usually compared to the Bose-Einstein condensation \cite{davis1995bose,jin1996collective}, the Fr\"{o}hlich condensate is a nonequilibrium phenomenon. 
Many researchers have investigated the corresponding quantum Hamiltonian \cite{wu1977bose,wu1978cooperative,wu1981frohlich}, the classification, the coherence and the phonon statistics \cite{reimers2009weak,zhang2019quantum} of the Fr\"{o}hlich condensate. Recently, the study has been extended to magnons \cite{demokritov2006bose,chumak2009bose}. 

Despite intense investigations, no unambiguous identification of a Fr\"{o}hlich condensate has been proved. 
Misochko et.al. reported a Fr\"{o}hlich condensate in a single crystal of bismuth but it is transient \cite{misochko2004transient}. Reimers et.al showed that coherent Fr\"{o}hlich condensate involves extremely large energies that are inaccessible in
a biological environment \cite{reimers2009weak}. Altfeder et.al reported the 
optical phonon condensate in heterostructures at room temperature, but their system is  at thermal equilibrium \cite{altfeder2017scanning}. Nardecchia et.al. reported a remarkable absorption feature around 0.314 THz in bovine serum albumin (BSA) when driven out of equilibrium by optical pumping, which might be a signal of Fr\"{o}hlich condensate \cite{nardecchia2018out}. Zhang et.al. suggested that Raman or infrared spectroscopy can be used to observe the Fr\"{o}hlich condensate in some modern proteins \cite{zhang2019quantum}.

\begin{figure}[b]
    \centering
    \includegraphics[width=\linewidth]{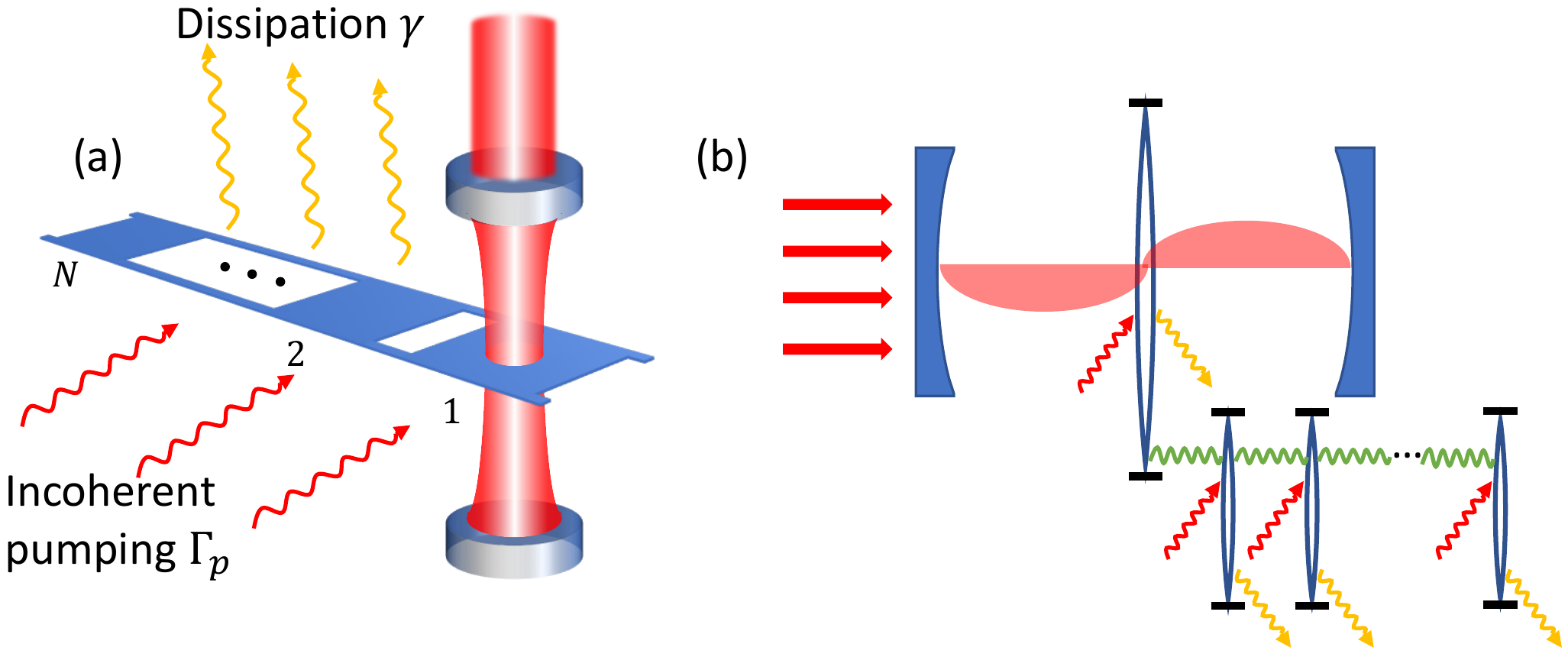}
    \caption{(a) Schematic of an optomechanical system: an optical cavity interacting with a one-dimensional array of membranes. (b) Simplified model. The dissipation is due to the contact of the system with environment. The incoherent phonon pumping is not necessary for our system to get Fr\"{o}hlich condensate, it is optional.}
    \label{fig:config}
\end{figure}

{\color{red}Currently, most of studies have been focused on biomolecules, the Fr\"{o}hlich condensate in non-biological systems is rarely investigated. Cavity optomechanics is a rapidly developing field exploring the interaction between electromagnetic radiation and micromechanical or nanomechanical motion \cite{kippenberg2008cavity,aspelmeyer2014cavity}. Due to the wide tuneability, feasible on-chip integration and hybridization with a variety of other quantum systems, optomechanical systems can be used in quantum transducers \cite{stannigel2010optomechanical,taylor2011laser,verhagen2012quantum,forsch2020microwave}, quantum storage \cite{zhang2003quantum,safavi2011electromagnetically,chang2011slowing,merklein2017chip}, high-precision measurement \cite{clerk2008back,o2010quantum,qvarfort2018gravimetry}, etc. These advantages inspire us to use optomechanical systems to realize a long-lived Fr\"{o}hlich condensate.}

In this paper, we propose for the first time that the Fr\"{o}hlich condensate of phonons can be realized in optomechanical systems, as opposed to the intensely investigated biological systems. To do so, we need to realize the two-phonon energy redistribution between different mechanical modes, {\color{red}which requires interactions between the optical field ($\hat{a}$) and the product of the displacements of different mechanical modes ($\hat{x}_i\hat{x}_j$) in the system. In experiment, a similar quadratic coupling $\hat{a}\hat{x}^2$ between optical field and single mechanical mode has been realized by placing a membrane in the middle of an optical cavity \cite{thompson2008strong,sankey2010strong}. Motivated by this, we extend the single membrane to an array of membranes to realize the interaction $\hat{a}\hat{x}_i\hat{x}_j$.} In our scheme, the optical field can be treated as a controllable reservoir \cite{tomadin2012reservoir,lee2018quantum}, which is quite different from the biological systems where the two-phonon processes are determined by the environment and cannot be controlled precisely. 

{\color{red}Our paper is organized as the following: In Section \ref{section:model}, we introduce the model for optomechanical system and obtain the rate equations for the average phonon numbers by the adiabatic elimination of optical field. Based on the rate equations, we investigate the dependence of Fr\"{o}hlich condensate on experimental parameters in Section \ref{section:frohlichcondensate}, \ref{section:experimentaldependence}. This includes the steady-state average phonon numbers at different detuning, optomechanical couplings and temperature, the critical optomechanical coupling strength required to achieve the Fr\"{o}hlich condensate, and the potential approaches to lower the critical coupling strength. In Section \ref{section:experfeasibility}, we discuss the feasibility of experimental implementation. Finally, we give outlook and conclusion in Section \ref{section:conclusion}. }

\section{methods and results}
\subsection{Hamiltonian and rate equations}
\label{section:model}
We consider a system consisting of a one-dimensional array of $N$ membranes with the first membrane positioned in the middle of an optical cavity, as shown in Fig. (\ref{fig:config}). The cavity is pumped by a driving laser. In this configuration, the cavity field couples to the square of the displacement of the first membranes \cite{thompson2008strong,sankey2010strong,liao2013photon,liao2014single}. Thus, the Hamiltonian for this system ($\hbar=1$) is{\color{red}
\begin{align}
    \hat{H}_0=&\omega_c \hat{a}^{\dagger}\hat{a}+E(\hat{a}^{\dagger}e^{-i\omega_dt}+\hat{a}e^{i\omega_dt})+G\hat{a}^{\dagger}\hat{a}\hat{x}_1^2\nonumber\\
    &+\sum_{j=1}^N\left(\frac{\hat{p}_j^2}{2m}+\frac{1}{2}k_j\hat{x}_j^2\right)-\sum_{j=1}^{N-1}k\hat{x}_j\hat{x}_{j+1},
\end{align}
where $\hat{a}$ ($\hat{a}^{\dagger}$) is the photon annihilation (creation) operator, $\omega_c$ is the frequency of the cavity mode, $\omega_d$ is the frequency of the coherent driving field with amplitude $E$. The optomechanical coupling strength $G=\omega_c^{\prime\prime}/2$ represents the frequency shift of the cavity mode induced by the square of displacement. For simplicity, the array of membranes is described as a chain of harmonic oscillators with the same mass $m$ and nearest-neighbor coupling $k$, and $\hat{p}_j$, $\hat{x}_j$, $k_j$ are the momentum operator, displacement operator and the internal spring constant of the $j$th membrane, respectively.

It is convenient to switch to a frame rotating at the driving frequency $\omega_d$. Applying the unitary transformation $\hat{U}=\exp{(-i\omega_d\hat{a}^{\dagger}\hat{a}t)}$ generates the new Hamiltonian
\begin{align}
     \hat{H}^{\prime}=&-\Delta \hat{a}^{\dagger}\hat{a}+E(\hat{a}^{\dagger}+\hat{a})+G\hat{a}^{\dagger}\hat{a}\hat{x}_1^2\nonumber\\
     &+\sum_{j=1}^N\left(\frac{\hat{p}_j^2}{2m}+\frac{1}{2}k_j\hat{x}_j^2\right)-\sum_{j=1}^{N-1}k\hat{x}_j\hat{x}_{j+1},
\end{align}
where $\Delta=\omega_d-\omega_c$ is the detuning. Splitting the cavity field into an average coherent amplitude and a fluctuating term $\hat{a}=\bar{\alpha}+\hat{d}$, and neglecting the higher order term $\hat{d}^{\dagger}\hat{d}$ with respect to $(\bar{\alpha}\hat{d}^{\dagger}+\bar{\alpha}^{\ast}\hat{d})$, we obtain the following Hamiltonian
\begin{align}
    \hat{H}=&-\Delta \hat{d}^{\dagger}\hat{d}+G(\bar{\alpha}\hat{d}^{\dagger}+\bar{\alpha}^{\ast}\hat{d})\hat{x}_1^2\nonumber\\
    &+\sum_{j=1}^N\left(\frac{\hat{p}_j^2}{2m}+\frac{1}{2}k_j^{\prime}\hat{x}_j^2\right)-\sum_{j=1}^{N-1}k\hat{x}_j\hat{x}_{j+1},\label{eq:Hamiltonian}
\end{align}
where $k^{\prime}_j=k_j+2G|\bar{\alpha}|^2\delta_{j,1}$ is the shifted internal spring constant. The coherent amplitude $\bar{\alpha}$ is given by 
\begin{align}
    \bar{\alpha}=\frac{-iE}{\kappa/2-i\Delta},
\end{align} 
where $\kappa$ is the decay rate of the optical mode. For simplicity, we assume that $k^{\prime}_j=k_0$. The mechanical part of Eq. (\ref{eq:Hamiltonian}) can be diagonalized with the well-known eigenfrequencies and eigenmodes
\begin{align}
    \omega_j^2&=\frac{k_0}{m}-\frac{2k}{m}\cos{\left(\frac{j\pi}{N+1}\right)},\\
    \bm{Y}&=M\bm{X},~\quad M_{i,j}=\sqrt{\frac{2}{N+1}}\sin{\left(\frac{ij\pi}{N+1}\right)},
\end{align}
where $\bm{Y}=[\hat{y}_1,\hat{y}_2,\dots,\hat{y}_N]^{T}$ and $\bm{X}=[\hat{x}_1,\hat{x}_2,\dots,\hat{x}_N]^{T}$ are the new and old basis, respectively. The Hamiltonian in the new basis is given by
\begin{align}
    \hat{H}=&-\Delta \hat{d}^{\dagger}\hat{d}+G(\bar{\alpha}\hat{d}^{\dagger}+\bar{\alpha}^{\ast}\hat{d})\left(\sum_{j=1}^N M_{j1}\hat{y}_j\right)^2\nonumber\\
    &+\sum_{j=1}^N\left(\frac{\hat{p}_{y,j}^2}{2m}+\frac{1}{2}m\omega_j^2\hat{y}_j^2\right).
\end{align}
Introducing the phonon creation ($\hat{b}_j^{\dagger}$) and annihilation ($\hat{b}_j$) operators, with
\begin{align}
    \hat{y}_j=\sqrt{\frac{\hbar}{2m\omega_j}}(\hat{b}_j^{\dagger}+\hat{b}_j),~ \hat{p}_{y,j}=-i\sqrt{\frac{\hbar m\omega_j}{2}}(\hat{b}_j^{\dagger}-\hat{b}_j),
\end{align}
we obtain the Hamiltonian}
\begin{align}
    \hat{H}&=-\Delta \hat{d}^{\dagger}\hat{d}+\sum_{j=1}^N\omega_j\hat{b}_j^{\dagger}\hat{b}_j+\frac{2g_0}{N+1}(\bar{\alpha} \hat{d}^{\dagger}+\bar{\alpha}^*\hat{d})\times\nonumber\\\times&\left[\sum_{j}U_{j,j}(\hat{b}_j+\hat{b}_j^{\dagger})^2+\sum_{i<j}2U_{i,j}(\hat{b}_i+\hat{b}_i^{\dagger})(\hat{b}_j+\hat{b}_j^{\dagger})\right], \label{eq:HamiltonianFinal}
\end{align}
where $g_0$ is the optomechanical coupling satisfying $g_0=Gx_0^2$, with $x_0=\sqrt{\hbar/(2m\omega_0)}$ and $\omega_0=\sqrt{k_0/m}$. The coefficient $U_{i,j}$ is given by
\begin{align}
    U_{i,j}=\frac{\omega_0}{{\sqrt{\omega_i\omega_j}}}\sin{\left(\frac{i\pi}{N+1}\right)}\sin{\left(\frac{j\pi}{N+1}\right)}.\label{eq:uij}
\end{align}

We are interested in studying the evolution and steady-state distribution of the phonon numbers at different modes. To include the effect of dissipation, we resort to the quantum master equation. The evolution of the system is governed by the quantum master equation,
\begin{align}
    \dot{\rho}=&-i[\hat{H},\rho]+\kappa\mathcal{D}[\hat{d}]\rho\nonumber\\
    &+\sum_{j=1}^N\left\{\gamma_j(1+\bar{n}_{j,\text{th}})\mathcal{D}[\hat{b}_j]\rho+\gamma_j\bar{n}_{j,\text{th}}\mathcal{D}[\hat{b}_j^{\dagger}]\rho\right\}, \label{eq:fullmaster}
\end{align}
where $\rho$ is the system density operator, $\hat{H}$ is given by Eq. (\ref{eq:HamiltonianFinal}), $\kappa$ and $\gamma_j$ are the decay rate of the optical and the $j$th mechanical modes, respectively. $\bar{n}_{j,\text{th}}$ is the thermal phonon population of the $j$th mechanical mode, and $\mathcal{D}[\hat{o}]\rho=\hat{o}\rho\hat{o}^{\dagger}-\frac{1}{2}(\hat{o}^{\dagger}\hat{o}\rho+\rho\hat{o}^{\dagger}\hat{o})$ is the Lindblad term. 
\begin{figure}[t]
    \centering
    \includegraphics[width=0.9\linewidth]{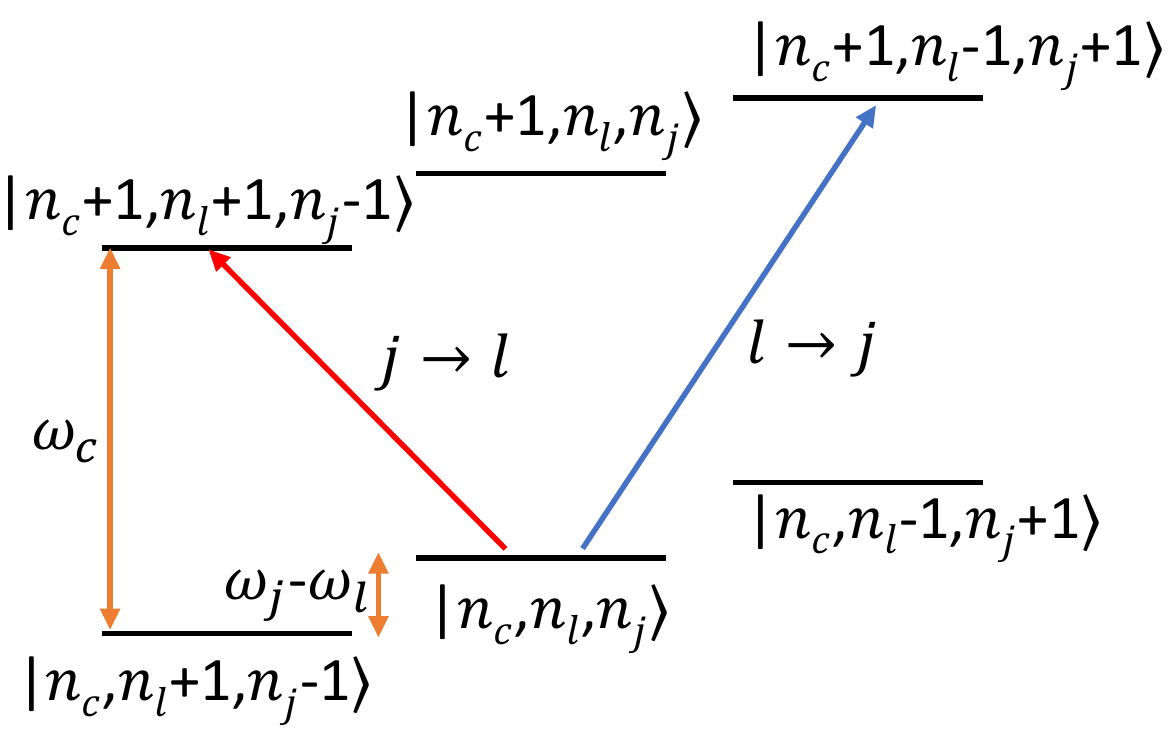}
    \caption{Transition processes between the $l$th mode and the $j$th mode in the case of $\omega_l<\omega_j$. Here $n_c$, $n_l$ and $n_j$ represent the photon number, phonon number at the $l$th mode and the $j$th mode, respectively. The red (left, $j\to l$) arrow describes the process such that one phonon at the $j$th mode is pumped to the $l$th mode by the red-detuned driving laser. The blue (right, $l\to j$) arrow describes the reverse process pumped by the blue-detuned driving laser.}
    \label{fig:transitionpicture}
\end{figure}Here we assume large optical dissipation $\kappa\gg\bar{n}_{j,\text{th}}\gamma_j$ and weak optomechanical coupling $\kappa,\omega_j \gg g_0|\bar{\alpha}|$. Hence, we can adiabatically eliminate the optical field by making the standard Born-Markov approximation \cite{carmichael2013statistical} and get the master equation for the reduced density operator $\rho_m$ of mechanical part. The rate equations for the average phonon numbers derived from the reduced master equation are given by (see Appendix \ref{appendix:rateequation} for the full derivation){\color{red}
\begin{widetext}
\begin{align}
   \langle\dot{\hat{n}}_l\rangle=& -\gamma_l(\langle \hat{n}_l\rangle-\bar{n}_{l,\text{th}})-2U_{l,l}^2\Gamma(2\omega_l)\langle \hat{n}_l(\hat{n}_l-1)\rangle+2U_{l,l}^2\Gamma(-2\omega_l)\langle(\hat{n}_l+1)(\hat{n}_l+2)\rangle\nonumber\\
&+\sum_{j\neq l}-4U_{j,l}^2\Gamma(\omega_l+\omega_j)\langle \hat{n}_l\hat{n}_j\rangle+4U_{j,l}^2\Gamma(-\omega_l-\omega_j)\langle(\hat{n}_l+1)(\hat{n}_j+1)\rangle\nonumber\\
&+\sum_{j\neq l}-4U_{j,l}^2\Gamma(\omega_l-\omega_j)\langle \hat{n}_l(\hat{n}_j+1)\rangle+4U_{j,l}^2\Gamma(\omega_j-\omega_l)\langle(\hat{n}_l+1)\hat{n}_j\rangle . \label{eq:fullrateequation}
\end{align}
\end{widetext}
where $\Gamma(\omega)$ is the transition rate and is given by  
\begin{align}
    \Gamma(\omega)=\frac{4g_0^2}{(N+1)^2}S_{nn}(\omega), \label{eq:gammafunction}
\end{align} 
with $S_{nn}(\omega)=\frac{4\kappa|\bar{\alpha}|^2}{4(\omega+\Delta)^2+\kappa^2}$ being the photon number spectral density.

The physical meaning of Eq. (\ref{eq:fullrateequation}) is explained as follows. In the first line, the first term describes the dissipation (one-phonon process) induced by the heat bath, the second and the third term describe the absorption and emission of two phonons at the $l$th mode, which are the well-known two-phonon sideband cooling and heating processes. The second line describes another kind of sideband cooling and heating processes, where instead of absorption or emission of two phonons at the same mode, the simultaneous absorption or emission of a phonon at the $l$th mode and the other phonon at the $j$th mode happened. The third line describes the energy-redistribution processes, where the first term describes the absorption of a phonon with energy $\hbar\omega_l$ in conjunction with emission of a phonon with energy $\hbar\omega_j$, and the second term describes the reverse processes, as illustrated in Fig. (\ref{fig:transitionpicture}). }
\begin{figure}[t]
    \centering
    \includegraphics[width=0.95\linewidth]{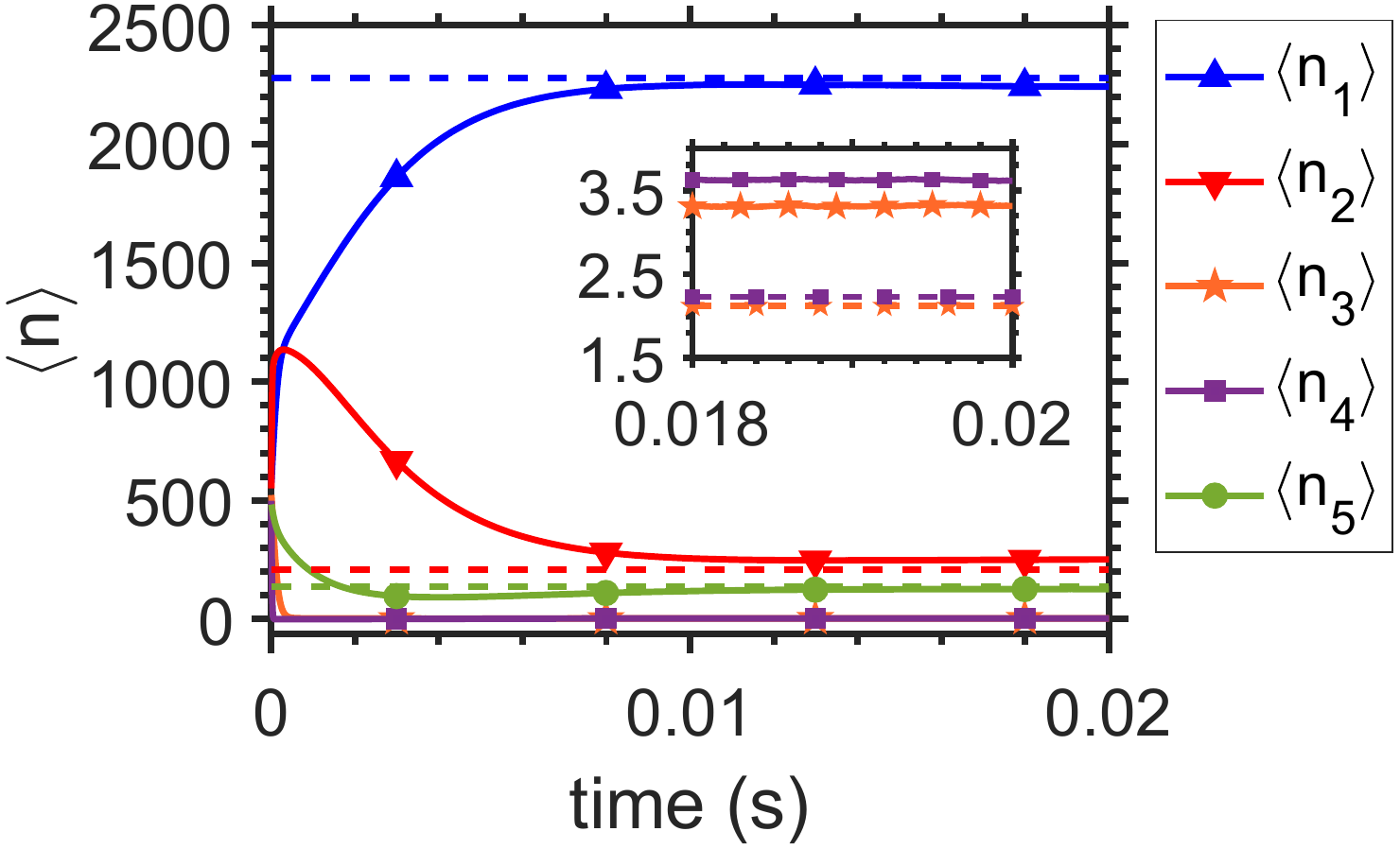}
    \caption{Comparison of average phonon numbers obtained from the MCWF method and decorrelation approximation in the case of $N=5$. The solid lines represent the evolution of the average phonon numbers obtained from MCWF method, and the dashed lines represent the steady-state phonon numbers obtained from decorrelation approximation. The zoomed-in inset shows the steady-state of the third and forth mode. The parameters are as follows: $T=400$ mK, $\omega_0=100$ MHz, $k/m=\omega_0^2/10$, $\kappa=1$ MHz, $\gamma=100$ Hz, $\Delta=-k/(m\omega_0)$,  $g_0|\bar{\alpha}|/\kappa=0.01$.}
    \label{fig:MCWF}
\end{figure}
\subsection{Fr\"{o}hlich condensate}
\label{section:frohlichcondensate}
{\color{red}Like in the Fr\"{o}hlich's model, the energy-redistribution processes are essential for realizing the condensate in our model (see Appendix \ref{appendix:comparison} for the comparison between Fr\"{o}hlich's original model and our model). In the next, we will focus on the energy-redistribution processes and neglect the unimportant sideband cooling and heating processes. Note that the spectral shape of rate $\Gamma(\omega)$ is Lorentzian that is centered at $-\Delta$ with linewidth $\kappa$.} Hence, the validity of neglecting sideband cooling and heating terms requires that the detuning satisfies $|\Delta|\approx|\omega_l-\omega_j|$, the frequency bandwidth of the mechancial system is narrow ($|\omega_l-\omega_j|\ll\omega_l,\omega_j$) so that $\Gamma(\pm2\omega_l) \ll \Gamma(\pm|\omega_j-\omega_l|)$, and the sideband cooling and heating processes are much weaker than the dissipation ($U_{l,l}^2\Gamma(\pm2\omega_l)\langle \hat{n}_l\rangle\ll \gamma_l$). All of these requirements are satisfied by the experimental parameters we used. In this case, Eq. (\ref{eq:fullrateequation}) can be simplified as
\begin{align}
    \langle\dot{\hat{n}}_l\rangle=&-\gamma_l(\langle \hat{n}_l\rangle-\bar{n}_{l,\text{th}})\nonumber\\
    &+\sum_{j\neq l}4U_{j,l}^2\left[\Gamma(\omega_j-\omega_l)\langle(\hat{n}_l+1)\hat{n}_j\rangle\right.\nonumber\\
&\left.-\Gamma(\omega_l-\omega_j)\langle \hat{n}_l(\hat{n}_j+1)\rangle\right].\label{eq:rateequation}
\end{align}
Summing Eq. (\ref{eq:rateequation}) over the mechanical mode $l$, we can remove the energy-redistribution terms:
\begin{align}
    \langle\dot{\hat{N}}_{\text{tot}}\rangle=-\sum_{l}\gamma_l(\langle \hat{n}_l\rangle-\bar{n}_{l,\text{th}}), \label{eq:totalnumbereq}
\end{align}
where $\langle \hat{N}_{\text{tot}}\rangle=\sum_{l}\langle \hat{n}_l\rangle$ is the total phonon number. We assume that $\gamma_l=\gamma$, then the evolution of total phonon number only depends on the dissipation processes. The steady state of $\langle \hat{N}_{\text{tot}}\rangle$ is given by $\langle \hat{N}_{\text{tot}}\rangle_{\text{ss}}=\sum_{l}\langle \bar{n}_{l,\text{th}}\rangle$, which is the summation of the thermal phonon populations at different mechanical modes.

The realization of Fr\"{o}hlich condensate is determined by the two-phonon-energy-redistribution processes. In the case of $\langle \hat{n}_j\rangle\gg 1$, the net transition rate from the $j$th mode to the lowest mode is approximately given by
\begin{align}
    \Gamma_{j\to 1}=4U_{j,1}^2[\Gamma(\omega_j-\omega_1)-\Gamma(\omega_1-\omega_j)]\langle \hat{n}_1 \hat{n}_j\rangle. \label{eq:transitiontolowest}
\end{align} 
From Eq. (\ref{eq:transitiontolowest}), we find $\Gamma_{j\to 1}>0$ for red detuning ($\Delta<0$) since $\Gamma(\omega)$ is Lorentzian centered at $-\Delta$. Similarly, the net transition rate from the $j$th mode to the highest mode is always positive ($\Gamma_{j\to N}>0$) for blue detuning ($\Delta>0$). Hence, phonons tend to concentrate at the lowest (highest) mode for red (blue) detuning. To realize Fr\"{o}hlich condensate at the lowest (highest) mode requires $\Gamma_{j\to 1}$ ($\Gamma_{j\to N}$) to be as large as possible. These rates can be increased by three approaches: (i) controlling the detuning $\Delta$; (ii) increasing the optomechanical couplings $g_0|\bar{\alpha}|$ by increasing the power of driving laser; (iii) increasing the product of phonon number $\langle \hat{n}_1\hat{n}_j\rangle$ by increasing the temperature or applying an external phonon pumping.

\begin{figure}[t]
    \centering
    \includegraphics[width=\linewidth]{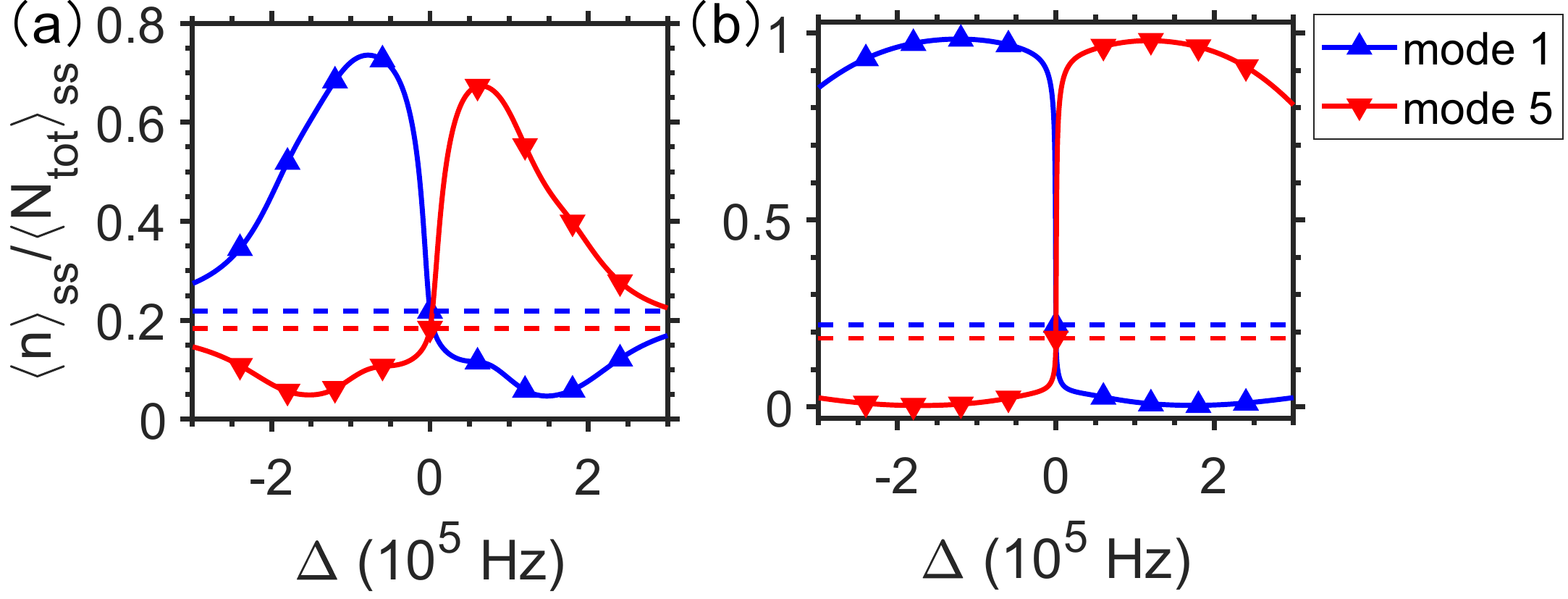}
    \caption{The steady-state ratio of phonon numbers with respect to detuning. (a) The optomechanical coupling $g_0|\bar{\alpha}|=10^{-5}\kappa$. (b) $g_0|\bar{\alpha}|=5 \times 10^{-5}\kappa$. $N=5$. The membrane and cavity parameters we use are close to those in Ref. \cite{thompson2008strong,sankey2010strong}. $\omega_0=1$ MHz, $\kappa=100$ kHz, $\gamma=0.1$ Hz, $T=300$ mK, the thermal phonon population $\bar{n}_{th}$ is at the order of $10^4$. The mechanical coupling is chosen as $k/m=\omega_0^2/10$. The upper (lower) dashed line denotes the ratio of phonon numbers at the lowest (highest) mode in thermal equilibrium. }
    \label{fig:nvsdelta}
\end{figure}

Note that the above analysis of the realization of Fr\"{o}hlich condensate is only based on the rate equations (Eq. (\ref{eq:fullrateequation}) and Eq. (\ref{eq:rateequation})), which are derived using adiabatic approximation. To do numerical integration, we need to introduce decorrelation approximation $\langle \hat{n}_l\hat{n}_j\rangle\approx\langle \hat{n}_l\rangle\langle \hat{n}_j\rangle$ to close the rate equations. The validity of decorrelation approximation in the estimation of the formation of Fr\"{o}hlich condensate is verified by using Monte Carlo wave-function (MCWF) method \cite{molmer1993monte} to integrate the master equation Eq. (\ref{eq:appendixreduced}). {\color{red}In Fig. \ref{fig:MCWF}, we show the evolution of average phonon numbers obtained from MCWF method by solid lines and the corresponding steady-state phonon numbers obtained from decorrelation approximation by dashed lines. We choose the average phonon numbers at $t=0.02$ s as the steady-state phonon numbers, and define the difference of the steady-state phonon number at the $i$th mode obtained from MCWF method and decorrelation approximation as $D_i=|\langle n_{i,\text{MC}}\rangle-\langle n_{\text{i,de}}\rangle|/\langle n_{\text{i,MC}}\rangle$. Using the parameters used in Fig. \ref{fig:MCWF}, we obtain $D_1=1.8 \%$, $D_2=18.4 \%$, $D_3=36.9 \%$, $D_4=37.8 \%$, $D_5=6.7 \%$. This suggests that the decorrelation approximation gives a good estimation of the steady-state average phonon number at the lowest mode. While the average phonon numbers at other modes obtained from decorrelation approximation can largely deviate from MCWF results, this deviation has very small effect on the estimation of the ratio $\langle n_1\rangle_{\text{ss}}/\langle N_{\text{tot}}\rangle_{\text{ss}}$ since the average phonon number at the lowest mode is dominant ($\langle n_{l>1}\rangle_{\text{ss}} \ll \langle n_1\rangle_{\text{ss}}$) and the total phonon number is almost constant ($\langle N_{\text{tot}}\rangle_{\text{ss}}\approx\sum_{l=1}^N\langle \bar{n}_{l,\text{th}}\rangle$). Hence, we can use the decorrelation approximation to estimate the value of $\langle n_1\rangle_{\text{ss}}/\langle N_{\text{tot}}\rangle_{\text{ss}}$, i.e., the formation of Frohlich condensate.
 In the next, we integrate Eq. (\ref{eq:fullrateequation}) under decorrelation approximation to investigate the dependence of Fr\"{o}hlich condensate on experimental parameters.}

\subsection{Dependence of condensation on experimental parameters}
\label{section:experimentaldependence}
In Fig. (\ref{fig:nvsdelta}), we show the steady-state ratio of average phonon numbers at the lowest and highest mode at different detuning, which agrees with our analysis of the transition rates Eq. (\ref{eq:transitiontolowest}) in Section \ref{section:frohlichcondensate}. The optimal detuning regime is $|\Delta|\sim k/(m\omega_0)=10^5$ Hz, which is about one half of the frequency band. If the detuning is far away from the optimal regime, the condensate no longer exists.

\begin{figure}[t]
    \centering
    \includegraphics[width=\linewidth]{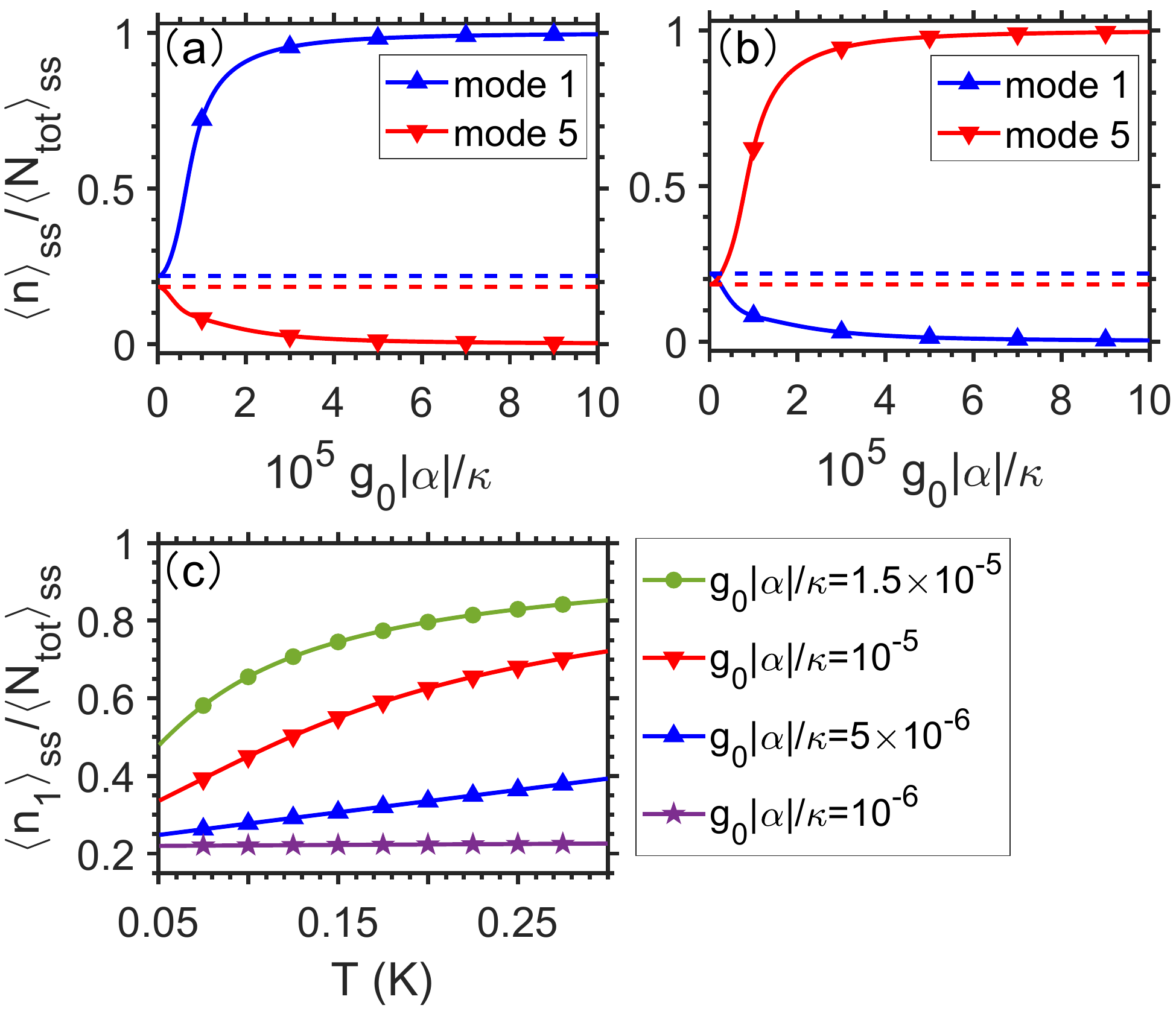}
    \caption{The first row: The steady-state ratio of phonon numbers with respect to optomechanical coupling. The temperature is fixed at $T=300$ mK. (a) Negative detuning $\Delta=-k/(m\omega_0)$. (b) Positive detuning $\Delta=k/(m\omega_0)$. The upper (lower) dashed line denotes the ratio of phonon numbers at the lowest (highest) mode in thermal equilibrium. The second row: (c) The steady-state ratio of phonon numbers with respect to temperature. $\Delta=-k/(m\omega_0)$. In (a)-(c), $N=5$, the cavity and membranes parameters are the same as those used in Fig. (\ref{fig:nvsdelta}). }
    \label{fig:nvsg0a}
\end{figure}

Besides the detuning, the realization of Fr\"{o}hlich condensate is largely affected by the optomechanical couplings and temperature. {\color{red}Fig. (\ref{fig:nvsg0a}) shows the dependence of the ratio $\langle n_1\rangle_{\text{ss}}/\langle N_{\text{tot}}\rangle_{\text{ss}}$ with respect to the optomechanical coupling and temperature. On the whole, phonons tend to concentrate at a single mode with the increase of optomechanical coupling and temperature. {\color{blue}In addition, the time needed to achieve the steady state is inversely proportional to the square of optomechanical coupling $g_0^2|\bar{\alpha}|^2$ (see Appendix \ref{appendix:timeevolution} for details).} Since the rate equations can not be solved analytically, here we give an order of magnitude estimate of the critical condition. The starting point is the rate equation for the lowest mode,
\begin{align}
     \langle\dot{\hat{n}}_1\rangle=&-\gamma(\langle \hat{n}_1\rangle-\bar{n}_{1,\text{th}})\nonumber\\
     &+\sum_{j=2}^N4U_{j,1}^2\left[\Gamma(\omega_j-\omega_1)(\langle \hat{n}_1\rangle+1)\langle \hat{n}_j\rangle\right.\nonumber\\
     &\left.-\Gamma(\omega_1-\omega_j)\langle \hat{n}_1\rangle(\langle \hat{n}_j\rangle+1)\right].\label{eq:rateequationlowestmode}
\end{align}
If the Fr\"{o}hlich condensate is achieved, we can assume $\langle \hat{n}_1\rangle_{\text{ss}}=a\langle \hat{N}_{\text{tot}}\rangle_{\text{ss}}$, where $a$ is a number close to 1. For large numbers of mechanical modes $N$, the total phonon number satisfies $\langle \hat{N}_{\text{tot}}\rangle_{\text{ss}}\gg \bar{n}_{1,th}$, then Eq. (\ref{eq:rateequationlowestmode}) at steady state can be simplified as  
\begin{align}
   &\gamma a\langle \hat{N}_{\text{tot}}\rangle_{\text{ss}}\nonumber\\
    &\approx 4a\langle \hat{N}_{\text{tot}}\rangle_{\text{ss}}\sum_{j=2}^NU_{j,1}^2\left[\Gamma(\omega_j-\omega_1)\right.
    \left.-\Gamma(\omega_1-\omega_j)\right]\langle \hat{n}_j\rangle_{\text{ss}} \label{eq:rateestimate}
\end{align}

\begin{figure}[t]
    \centering
    \includegraphics[width=\linewidth]{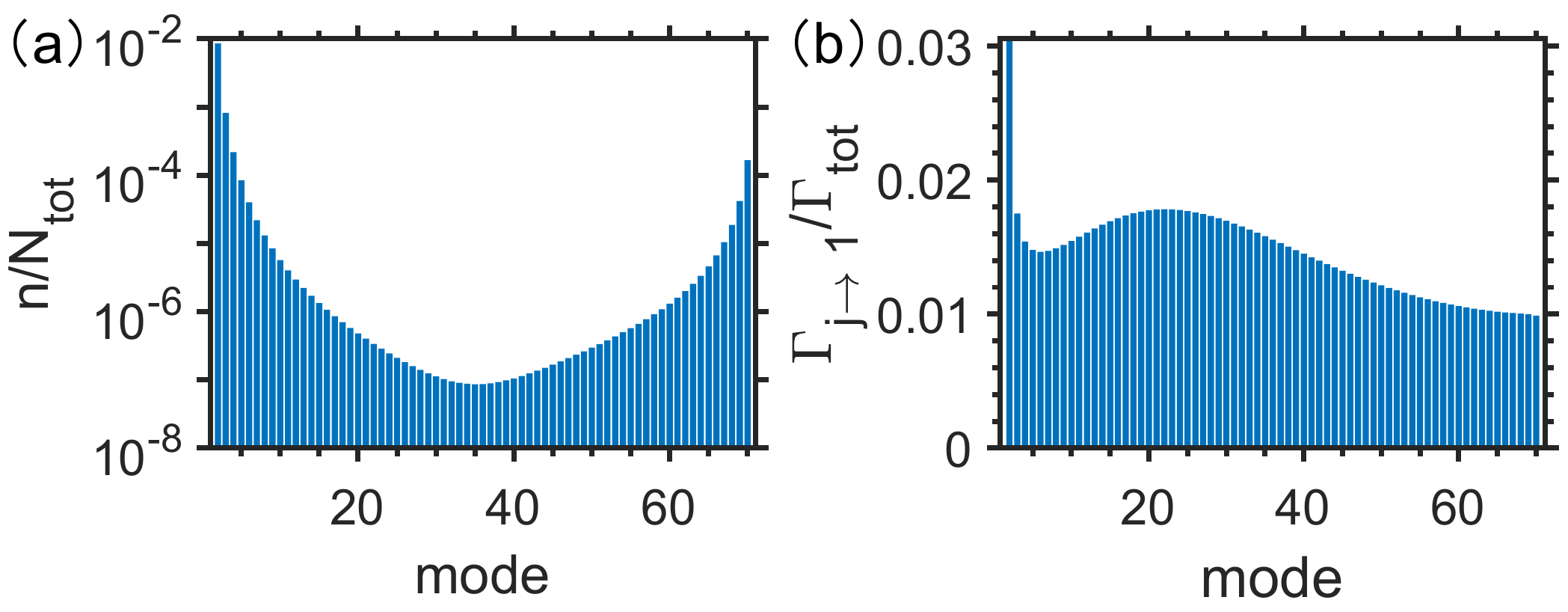}
    \caption{ The steady state of a Fr\"{o}hlich condensate with $\langle \hat{n}_1\rangle/\langle \hat{N}_{\text{tot}}\rangle=0.99$. (a) Phonon distribution in $j\ge 2$ modes. (b) The net transition rate from the $j$th mode to the lowest mode. $N=70$, $g_0|\bar{\alpha}|=0.06\kappa$, $\Delta=-k/(m\omega_0)$, the other cavity and membranes parameters are the same as those used in Fig. (\ref{fig:nvsdelta}).}
    \label{fig:phonondistribution}
\end{figure}

\begin{figure*}[t]
    \centering
    \includegraphics[width=0.8\linewidth]{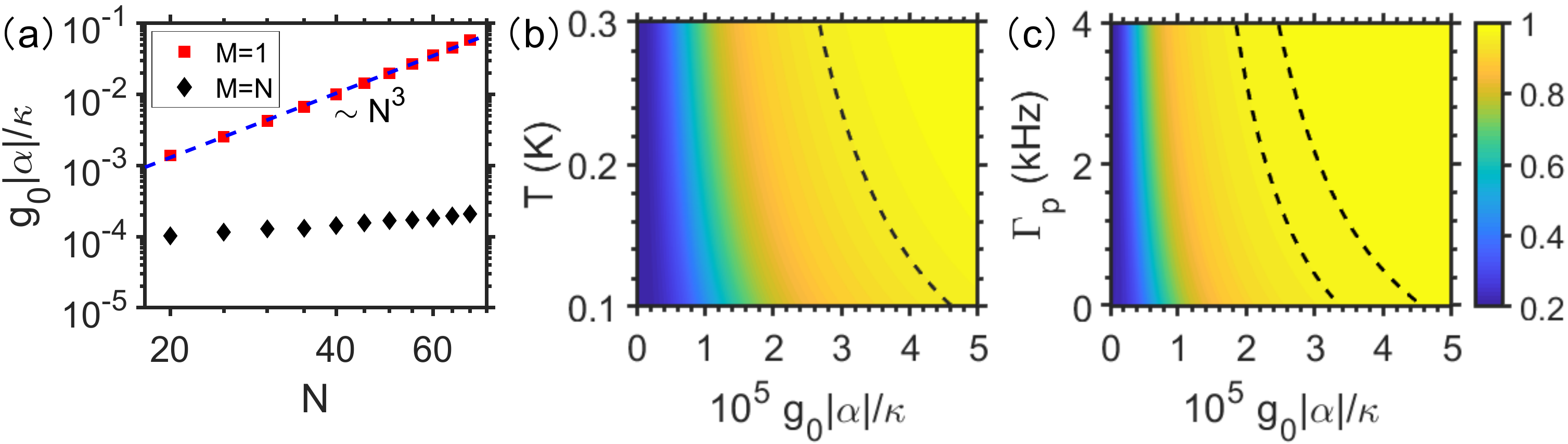}
    \caption{ Critical condition of Fr\"{o}hlich condensate. (a) The minimum value of optomechanical coupling required to realize $a=0.99$ with respect to the number of membranes. The red squares represent the case of a single cavity and the balck diamonds represent the case of $N$ cavities. (b) The contour of $\langle \hat{n}_1\rangle/\langle \hat{N}_{\text{tot}}\rangle$ in steady state with respect to $g_0|\bar{\alpha}|$ and temperature in the case of $N=5$. The dashed line fitted by Eq. (\ref{eq:criticalcondition}) agrees well with the contour lines $\langle \hat{n}_1\rangle/\langle \hat{N}_{\text{tot}}\rangle=0.945$. (c) The contour of $\langle \hat{n}_1\rangle/\langle \hat{N}_{\text{tot}}\rangle$ in steady state with respect to $g_0|\bar{\alpha}|$ and external phonon pumping in the case of $N=5$. The temperature $T=300$ mK is fixed. The dashed lines are fitted by Eq. (\ref{eq:criticalcondition}) with $\bar{n}_{\text{th}}$ replaced by $\frac{\Gamma_p}{\gamma}+\bar{n}_{\text{th}}$, which agree well the contour line $\langle \hat{n}_1\rangle/\langle \hat{N}_{\text{tot}}\rangle=0.966$ and $\langle \hat{n}_1\rangle/\langle \hat{N}_{\text{tot}}\rangle=0.98$. In (a) - (c), $\Delta=-k/(m\omega_0)$, the cavity and membrane parameters are fixed and are the same as those used in Fig. (\ref{fig:nvsdelta}).}
    \label{fig:contour}
\end{figure*}
To estimate the order of magnitude of the summation on the right hand-side of Eq. (\ref{eq:rateestimate}), we first plot the steady-state average phonon distribution of the Fr\"{o}hlich condensate with $a=0.99$ and the corresponding net transition rate from the $j$th mode to the lowest mode $\Gamma_{j\to 1}$ in the case of $N=70$, as shown in Fig. (\ref{fig:phonondistribution}). It can be seen that while the rest of phonons mainly occupy the second mode, the net transition rates from $j\ge2$ modes to the lowest mode are at the same order of magnitude. Based on this observation, we assume the phonon number at the second mode is approximately $(1-a)\langle \hat{N}_{\text{tot}}\rangle_{\text{ss}}$ and consider the transition from  the second mode to the lowest mode
\begin{align}
    &\Gamma_{2\to 1}\approx 4a(1-a)\langle \hat{N}_{\text{tot}}\rangle^2_{\text{ss}} U_{2,1}^2\left[\Gamma(\omega_2-\omega_1)\right.\left.-\Gamma(\omega_1-\omega_2)\right].
\end{align}
In the large $N$ limit, the frequency difference between the first two modes satisfies $\omega_2-\omega_1\sim k/(m\omega_0N^2)\ll |\Delta|,\kappa$. Expanding $\Gamma(\omega_2-\omega_1)-\Gamma(\omega_1-\omega_2)$ to the first order of $\omega_2-\omega_1$ and making use of the formula of $U_{2,1}^2$, we obtain the order of magnitude estimate of the net transition rate $\Gamma_{2\to1}$ 
\begin{align}
    \Gamma_{2\to 1}\sim \frac{6000\pi^6a(1-a)k\kappa|\Delta| g_0^2|\bar{\alpha}|^2}{m\omega_0(\kappa^2+4\Delta^2)^2N^8}\langle \hat{N}_{\text{tot}}\rangle^2_{\text{ss}}.
\end{align}
The total transition rate from high frequency modes to the lowest one is approximately given by $N\Gamma_{2\to1}/2$, which should be equal to the dissipation rate $\gamma a\langle \hat{N}_{\text{tot}}\rangle_{\text{ss}}$ at the steady state. Then we can obtain the critical condition required to achieve Fr\"{o}hlich condensate}
\begin{align}
    \frac{3000\pi^6(1-a)k\kappa|\Delta| g_0^2|\bar{\alpha}|^2\bar{n}_{\text{th}}}{m\omega_0\gamma(\kappa^2+4\Delta^2)^2}\sim N^6, \label{eq:criticalcondition}
\end{align}
where $\bar{n}_{\text{th}}=\frac{1}{N}\sum_{l}\langle \bar{n}_{l,\text{th}}\rangle$ is the mean thermal phonon number. In the derivation of Eq. (\ref{eq:criticalcondition}) we have chosen $\Delta=-k/(m\omega_0)$ and made use of $\langle \hat{N}_{\text{tot}}\rangle_{\text{ss}}=\sum_{l}\langle \bar{n}_{l,\text{th}}\rangle$.  In the limit of high temperature $k_BT\gg \hbar\omega_0$, the mean thermal phonon number is approximately given by $\bar{n}_{\text{th}}\approx \frac{k_BT}{\hbar\omega_0}$, where $k_B$ is the Boltzmann constant.

Eq. (\ref{eq:criticalcondition}) shows that the product of $g_0^2|\bar{\alpha}|^2$ and temperature $T$ is proportional to $N^{6}$. To verify our estimation, we plot the minimum value of $g_0|\bar{\alpha}|$ required to achieve a condensation with $a=0.99$ with all other parameters fixed, as shown by the red squares in Fig. (\ref{fig:contour}a). It is seen that $g_0|\bar{\alpha}|$ is proportional to $N^{3}$, which agrees well with our estimation. Fig. (\ref{fig:contour}b) shows the contour of $\langle \hat{n}_1\rangle/\langle \hat{N}_{\text{tot}}\rangle$ in steady state with respect to $g_0|\bar{\alpha}|$ and temperature. The dashed line is of the form $g_0^2|\bar{\alpha}|^2T=\text{const}$, which agrees well with the contour line of $\langle \hat{n}_1\rangle/\langle \hat{N}_{\text{tot}}\rangle$. Unlike the Fr\"{o}hlich's model in which the critical pumping rate is proportional to $1/N$ \cite{zhang2019quantum}, in our model the critical product of $g_0^2|\bar{\alpha}|^2$ and temperature $T$ is proportional to $N^6$. This makes the realization of Fr\"{o}hlich condensate in large $N$-membrane systems challenging. The $N^6$ scaling mainly comes from the coefficient $1/(N+1)^2$ of the rate function $\Gamma(\omega)$ and the sine functions in $U_{j,l}$. 

To lower the requirements of implementing the Fr\"{o}hlich condensate, we can introduce multiple cavities along the array of membranes, with each cavity coupled to the corresponding membrane at the cavity position. Using adiabatic approximation, we can get rate equations in the same form as the rate equations for a single cavity, except that the coefficient $U_{i,j}^2$ is replaced by the effective one $\tilde{U}_{i,j}^2$ (see Appendix \ref{appendix:multiplecavities} for details). We denote the number of cavities as $M$. In the case of $M=N$, i.e., every membrane is coupled to a corresponding cavity, we can get a simple expression for the effective coefficient, 
\begin{align}
    \tilde{U}_{i,j}^2=\frac{(N+1)\omega_0^2}{\omega_i\omega_j}\times\begin{cases}
    \frac{3}{8}, & \text{for $i=j$}.\\
    \frac{1}{4}, & \text{for $i\neq j$}.
  \end{cases}
    \label{eq:effectiveuij}
\end{align}
Depending on the value of $i$ and $j$, the effective coefficient $\tilde{U}_{i,j}^2$ can be greater than the coefficient $U_{i,j}^2$ (Eq. (\ref{eq:uij})) by a factor between $N$ and $N^5/\pi^4$ in the large $N$ limit. Hence, the required optomechanical coupling or temperature is orders of magnitude lower than that of the single cavity case, as shown by the black diamonds in Fig. (\ref{fig:contour}a).  
\begin{table*}[t]
\begin{tabular}{|c|c|c|c|c|c|c|c|c|c|c|c|}
\hline
N & Membrane size                                                                            & $\omega_0/2\pi$ & Mass   & $Q_m$           & $\lambda$ & $\kappa$ & $G/2\pi$                & $k/m$                  & $\Delta$                 & T      & $P_{\text{in}}$ \\ \hline
5 & \begin{tabular}[c]{@{}c@{}}1 mm $\times$ 1 mm \\ $\times$ 50 nm\end{tabular}             & 134 kHz         & 40 ng  & $1.2\times10^7$ & 1064 nm   & 300 kHz  & 15 MHz$\cdot$nm$^{-2}$ & $\frac{\omega_0^2}{10}$ & $\frac{-k}{m\omega_0}$ & 300 mK & 5 $\mu$W        \\ \hline
5 & \begin{tabular}[c]{@{}c@{}}100 $\mu$m $\times$ 100 $\mu$m \\ $\times$ 50 nm\end{tabular} & 1.34 MHz        & 400 pg & $1.2\times10^6$ & 1064 nm   & 300 kHz  & 15 MHz$\cdot$nm$^{-2}$ & $\frac{\omega_0^2}{10}$ & $\frac{-k}{m\omega_0}$ & 300 mK & 7 mW            \\ \hline
\end{tabular} 
\caption{Membrane and cavity parameters and the estimated power of driving laser.}
\label{tab:feasibility}
\end{table*}

To further lower the requireed optomechanical coupling, we can also introduce the incoherent phonon pumping used in Fr\"{o}hlich's original model to our system. The incoherent pumping can increase the phonon numbers in the system. If we assume the pumping rate is $\Gamma_p$, then the total phonon number in the system is $\langle N_{\text{tot}}\rangle=\frac{N\Gamma_p}{\gamma}+N\bar{n}_{\text{th}}$. Hence we can introduce the effective temperature of the system $T_{\text{eff}}\approx T+\frac{\hbar\omega_0\Gamma_p}{k_B\gamma}$. The extra term $\frac{\hbar\omega_0\Gamma_p}{k_B\gamma}$ in the effective temperature can lower the requirement of temperature and optomechanical coupling, as shown in Fig. (\ref{fig:contour}c).

\subsection{Experimental feasibility}
\label{section:experfeasibility}
In the above section, we have shown that the Fr\"{o}hlich condensate can be achieved in optomechanical systems via numerical simulations. 
Here we estimate the experimental feasibility by using existing systems. The experimental parameters are shown in Table \ref{tab:feasibility}.

We first consider the membrane and cavity parameters of Ref. \cite{sankey2010strong}. The Si$_3$N$_4$ membrane used in Ref. \cite{sankey2010strong} has the size of 1 mm $\times$ 1 mm $\times$ 50 nm with resonance at $\omega_0=2\pi\times134$ kHz and mass $m=40$ ng. The cavity allows the existence of many transverse electromagnetic (TEM) modes. The TEM$_{00}$ mode of the cavity has a decay rate $\kappa=300$ kHz. The cavity is excited by a laser with wavelength $\lambda=1064$ nm. The coupling between the mechanical mode and TEM$_{00}$ mode at the TEM$_{02}$–TEM$_{00}$ crossing can be as large as $G=\omega_c^{\prime\prime}/2=2\pi\times15$ MHz$\cdot$nm$^{-2}$, which corresponds to $g_0=Gx_0^2=1.5\times10^{-4}$ Hz. At $T=300$ mK, the Q factor of the membrane is $Q_m=1.2\times10^7$, which corresponds to the mechanical decay rate $\gamma=0.07$ Hz. We assume the coupling between membranes is $k/m=\omega_0^2/10$, the detuning is $\Delta=-k/(m\omega_0)$ and the number of membranes is $N=5$. Using Eq. (\ref{eq:criticalcondition}), we find the estimated coherent amplitude $|\bar{\alpha}|$ of the cavity field required to achieve $\langle n_1\rangle_{\text{ss}}/\langle N_{\text{tot}}\rangle_{\text{ss}}=0.99$ is $|\bar{\alpha}|=1.6\times10^4$. Using the formula $\bar{\alpha}=-iE/(\kappa/2-i\Delta)$ and $E=\sqrt{\frac{\kappa P_{\text{in}}}{\hbar\omega_d}}$, we can calculate the required laser power, which is about $P_{\text{in}}\approx 5$ $\mu$W. This is feasible in nowadays lab.

{\color{red}The optomechanical system we used is scalable. If we decrease the size of membrane to 100 $\mu$m $\times$ 100 $\mu$m $\times$ 50 nm, the resonance frequency increases to $\omega_0=2\pi\times1.34$ MHz, the mass decreases to $m=400$ pg. Usually, the Q factor will decrease as the size decreases \cite{verbridge2006high,barton2011high}. We assume the Q factor of membranes decreases to $Q_m=1.2\times10^6$ (namely changes 10 times) and the corresponding mechanical decay rate becomes $\gamma=7$ Hz. We further assume the coupling $G$ and the optical decay rate $\kappa$ do not change, the coupling between membranes and the detuning still follow $k/m=\omega_0^2/10$ and $\Delta=-k/(m\omega_0)$. In this case, the required laser power is $P_{\text{in}}\approx 7$ mW. This power is achievable in optomechanical experiments \cite{colombano2019synchronization}. Of course, one can also assume the Q value does not change \cite{sankey2010strong}, then the laser power required should be $P_{\text{in}}\approx 700 $ $\mu$W, which is also available in experiments.

We can also estimate the vibrational amplitude of the membranes when the Fr\"{o}hlich condensate is achieved. The parameters of 1 mm $\times$ 1 mm $\times$ 50 nm Si$_3$N$_4$ membrane are used. The temperature is $T=300$ mK. At the state of Fr\"{o}hlich condensate, the phonons mainly occupy the lowest mode, the phonon number of the lowest mode is approximately $\frac{Nk_BT}{\hbar\omega_0}$, then the total energy concentrated at the lowest mode is $\frac{Nk_BT}{\hbar\omega_0}\hbar\omega_1\approx Nk_BT$. The amplitude of the membranes can be estimated by the relationship $m\omega_1^2\langle x^2\rangle=Nk_BT$, which gives the vibrational amplitude $A\approx0.8$ pm for $N=5$. This amplitude is detectable in nowadays experiments \cite{thompson2008strong,singh2014optomechanical}. The amplitude can be further increased by applying external phonon pumping or increasing the temperature, as we mentioned in Sec. \ref{section:experimentaldependence}.}

\section{Discussion and conclusion}
\label{section:conclusion}

In this paper, we have discussed the realization of Fr\"{o}hlich condensate in an array of membranes quadratically coupled with a cavity. Besides this setup, we can also consider quadratically coupling the multiple modes of a single membranes to the cavity. In experiment, the linear interaction between multiple mechanical modes of a membrane and an optical mode has been reported \cite{nielsen2017multimode}. The realization of quadratic interaction in a similar platform is promising. With a much larger quadratic coupling strength $g_0|\bar{\alpha}|/\kappa\approx 1$ \cite{murch2008observation,purdy2010tunable}, the optomechanical interaction between ultracold atomic gas and cavity can be a candidate to realize Fr\"{o}hlich condensate as well. 

In conclusion, we have proposed that the Fr\"{o}hlich condensate can be realized in an array of mechanical membranes coupled to an optical cavity. Compared to the biological systems studied before, the optomechanical system can be well controlled by the driving laser. The numerical results show that the optimal detuning of driving laser is around $|\Delta|\sim k/(m\omega_0)$. The optomechanical coupling strength and temperature required to achieve Fr\"{o}hlich condensate satisfy the relationship $g_0|\bar{\alpha}|^2T\propto N^6$. The possible solution to lower the requirement is introducing multiple cavities and incoherent phonon pumping to the system .

Fr\"{o}hlich condensates in optomechanical systems have potential applications in the energy conversion and transfer, frequency filter, heat control \cite{seif2018thermal,yang2020phonon}, etc. By switching the role of phonon and photon, we may realize the Fr\"{o}hlich-like photon condensate, which may be used in acoustic-induced-optical lasing. Since most phonons occupy the lowest mode in the Fr\"{o}hlich condensate, we may realize the efficient multimode cooling of mechanical oscillators by adding another laser to cool the lowest mode.

\begin{acknowledgments}
This work utilized resources from the University of Colorado Boulder Research Computing Group, which is supported by the National Science Foundation (awards ACI-1532235 and ACI-1532236), the University of Colorado Boulder, and Colorado State University. 
\end{acknowledgments}
\appendix

\section{Quantum master equations and rate equations for phonon numbers}
\label{appendix:rateequation}

To derive the effective quantum master equations involving only the mechanical degrees of freedom, we can adiabatically eliminate the optical degrees of freedom. The validity of adiabatic elimination relies on two approximations: (i) the optical decay rate is much larger than the mechanical dissipation rate, i.e. the optical processes are much faster than mechanical processes. (ii) the optomechanical coupling is weak. The master equation for the system density operator $\rho$ is given by
\begin{align}
 &\dot{\rho}=\mathcal{L}_o\rho+\mathcal{L}_m\rho+\mathcal{L}_{om}\rho\\
    &\mathcal{L}_o=-i[-\Delta \hat{d}^{\dagger}\hat{d},\cdot]+\kappa\mathcal{D}[\hat{d}]\\
    &\mathcal{L}_m=\sum_{j=1}^N\left\{-i[\omega_j\hat{b}_j^{\dagger}\hat{b}_j,\cdot]\right.\nonumber\\
    &~~~~~~~~\left.+\gamma_j(1+\bar{n}_{j,\text{th}})\mathcal{D}[\hat{b}_j]+\gamma_j\bar{n}_{j,\text{th}}\mathcal{D}[\hat{b}_j^{\dagger}]\right\}\\
    &\mathcal{L}_{om}=-i[\hat{H}_{om},\cdot].
\end{align}
where $\hat{H}_{om}$ is the optomechanical coupling term in Eq. (\ref{eq:HamiltonianFinal}). In the interaction picture, we transform the density operator as 
\begin{align}
    \rho^I(t)=e^{-(\mathcal{L}_o+\mathcal{L}_m)t}\rho.
\end{align}
Then the master equation is transformed as
\begin{align}
    &\dot{\rho}^I=\mathcal{L}^I_{om}(t)\rho^I, \label{eq:interactionmaster}
\end{align}
with
\begin{align}
    \mathcal{L}^I_{om}(t)=e^{-(\mathcal{L}_o+\mathcal{L}_m)t}\mathcal{L}_{om}e^{(\mathcal{L}_o+\mathcal{L}_m)t}.
\end{align}
The formal solution of Eq. (\ref{eq:interactionmaster}) is given by
\begin{align}
    \rho^I(t)=\rho^I(0)+\int_0^tdt_1\mathcal{L}^I_{om}(t_1)\rho^I(t_1).
\end{align}
Substituting the formal solution back into Eq. (\ref{eq:interactionmaster}) gives
\begin{align}
    \dot{\rho}^I(t)=\mathcal{L}^I_{om}(t)\rho^I(0)+\int_0^tdt_1\mathcal{L}^I_{om}(t)\mathcal{L}^I_{om}(t_1)\rho^I(t_1).
\end{align}
We assume that no correlations exist between the optical and mechanical field at $t=0$. Then $\rho^I(0)=\rho(0)=\rho_o(0)\otimes\rho_m(0)$. Taking the partial trace over the optical degrees of freedom results in
\begin{align}
     \dot{\rho}^I_m(t)=\int_0^tdt_1\text{Tr}_o\left\{\mathcal{L}^I_{om}(t)\mathcal{L}^I_{om}(t_1)\rho^I(t_1)\right\},\label{masteroptical}
\end{align}
where, for simplicity, we have eliminated the term $\text{Tr}_o\left\{\mathcal{L}^I_{om}(t)\rho^I(0)\right\}$ with the assumption $\text{Tr}_o\left\{\hat{d}\rho_o(0)\right\}=\text{Tr}_o\left\{\hat{d}^{\dagger}\rho_o(0)\right\}=0$. Since the optomechanical coupling is weak, the density operator can be written as
\begin{align}
    \rho^I(t)=\rho^I_o(t)\otimes\rho^I_m(t)+O(\hat{H}_{om}).
\end{align}
Now we make the standard Born-Markov approximation, the master equation for the reduced density operator is
\begin{align}
    \dot{\rho}^I_m(t)=\int_0^tdt_1\text{Tr}_o\left\{\mathcal{L}^I_{om}(t)\mathcal{L}^I_{om}(t_1)\rho^I_o(t)\otimes\rho^I_m(t)\right\}. \label{eq:bornmarkov}
\end{align}
The form of $\hat{H}_{om}$ is
\begin{align}
    \hat{H}_{om}=\epsilon\hat{A}\hat{B},
    \label{eq:hom}
\end{align}
with
\begin{align}
    \epsilon&=\frac{2g_0}{N+1}\nonumber\\
    \hat{A}&=(\bar{\alpha} \hat{d}^{\dagger}+\bar{\alpha}^*\hat{d})\nonumber\\
    \hat{B}&=\sum_{i}U_{i,i}(\hat{b}_i+\hat{b}_i^{\dagger})^2+\sum_{i<j}2U_{i,j}(\hat{b}_i+\hat{b}_i^{\dagger})(\hat{b}_j+\hat{b}_j^{\dagger}).
    \label{eq:homdetail}
\end{align}
Plugging Eqs. (\ref{eq:hom})-(\ref{eq:homdetail}) into Eq. (\ref{eq:bornmarkov}), we obtain
\begin{align}
    &\dot{\rho}^I_m(t)\nonumber\\
    &=-\epsilon^2\int_0^tdt_1\left\{(\hat{B}\cdot)_t(\hat{B}\cdot)_{t_1}\rho^I_m(t)\text{Tr}_o\left[(\hat{A}\cdot)_t(\hat{A}\cdot)_{t_1}\rho^I_o(t)\right]\right.\nonumber\\
    &~~~\left.-(\hat{B}\cdot)_{t}(\cdot\hat{B})_{t_1}\rho^I_m(t)\text{Tr}_o\left[(\hat{A}\cdot)_{t}(\cdot\hat{A})_{t_1}\rho^I_o(t)\right]\right.\nonumber\\
    &~~~\left.-(\cdot\hat{B})_t(\hat{B}\cdot)_{t_1}\rho^I_m(t)\text{Tr}_o\left[(\cdot\hat{A})_t(\hat{A}\cdot)_{t_1}\rho^I_o(t)\right]\right.\nonumber\\
    &~~~\left.+(\cdot\hat{B})_t(\cdot\hat{B})_{t_1}\rho^I_m(t)\text{Tr}_o\left[(\cdot\hat{A})_t(\cdot\hat{A})_{t_1}\rho^I_o(t)\right]\right\},\label{eq:masterAB}
\end{align}
where $(\hat{O}\cdot)_t=e^{-(\mathcal{L}_o+\mathcal{L}_m)t}(\hat{O}\cdot)e^{(\mathcal{L}_o+\mathcal{L}_m)t}$ and $(\cdot\hat{O})_t=e^{-(\mathcal{L}_o+\mathcal{L}_m)t}(\cdot\hat{O})e^{(\mathcal{L}_o+\mathcal{L}_m)t}$ are the operators in the interaction picture. The dot represents where the density operator is placed. Taking the time derivative on both sides, we obtain the equation satisfied by the optical operator $(\hat{d}\cdot)_t$, $(\cdot\hat{d})_t$:
\begin{align}
    \frac{d(\hat{d}\cdot)_t}{dt}=\left[(\hat{d}\cdot)_t,\mathcal{L}_o\right],\quad \frac{d(\cdot\hat{d})_t}{dt}=\left[(\cdot\hat{d})_t,\mathcal{L}_o\right].\label{dbdt}
\end{align}
From Eq. (\ref{dbdt}) we can obtain the explicit solutions
\begin{align}
    (\hat{d}\cdot)_t&=e^{(i\Delta-\frac{\kappa}{2})t}(\hat{d}\cdot),\nonumber\\
    (\hat{d}^{\dagger}\cdot)_t&=e^{-(i\Delta-\frac{\kappa}{2})t}(\hat{d}^{\dagger}\cdot)-e^{-(i\Delta+\frac{\kappa}{2})t}\left(e^{\kappa t}-1\right)(\cdot \hat{d}^{\dagger}). \label{eq:dt}
\end{align}
For the mechanical operators $(\hat{b}_j\cdot)_t$ and $(\hat{b}^{\dagger}_j\cdot)_t$ in the interaction picture, we have
\begin{align}
    (\hat{b}_j\cdot)_t=e^{-i\omega_jt}(\hat{b}_j\cdot),\quad
    (\hat{b}^{\dagger}_j\cdot)_t=e^{i\omega_jt}(\hat{b}^{\dagger}_j\cdot), \label{eq:bt}
\end{align}
with the assumption that $\omega_j\gg\gamma_j$. We have assumed that $\kappa\gg\bar{n}_{j,\text{th}}\gamma_j, g_0|\bar{\alpha}|$, i.e., the time scale of mechanical dynamics is much larger than that of optical dynamics. Then $\rho^I_o(t)\approx\rho^I_o(\infty)$. Making use of Eq. (\ref{eq:dt}), we get the explicit expressions for the partial traces in Eq. (\ref{eq:masterAB})
\begin{align}
    \text{Tr}_o\left[(\hat{A}\cdot)_t(\hat{A}\cdot)_{t_1}\rho^I_o(t)\right]&=|\bar{\alpha}|^2e^{(i\Delta-\frac{\kappa}{2})(t-t_1)}\nonumber\\
    \text{Tr}_o\left[(\cdot\hat{A})_t(\hat{A}\cdot)_{t_1}\rho^I_o(t)\right]&=|\bar{\alpha}|^2e^{(i\Delta-\frac{\kappa}{2})(t-t_1)}\nonumber\\
    \text{Tr}_o\left[(\hat{A}\cdot)_{t}(\cdot\hat{A})_{t_1}\rho^I_o(t)\right]&=|\bar{\alpha}|^2e^{-(i\Delta+\frac{\kappa}{2})(t-t_1)}\nonumber\\
    \text{Tr}_o\left[(\cdot\hat{A})_t(\cdot\hat{A})_{t_1}\rho^I_o(t)\right]&=|\bar{\alpha}|^2e^{-(i\Delta+\frac{\kappa}{2})(t-t_1)}.\label{eq:correl}
\end{align}
From Eqs. (\ref{eq:bt})-(\ref{eq:correl}), and neglecting the fast-oscillating terms, we can simplify the master equation Eq. (\ref{eq:masterAB}):
\begin{figure}[t]
    \centering
    \includegraphics[width=\linewidth]{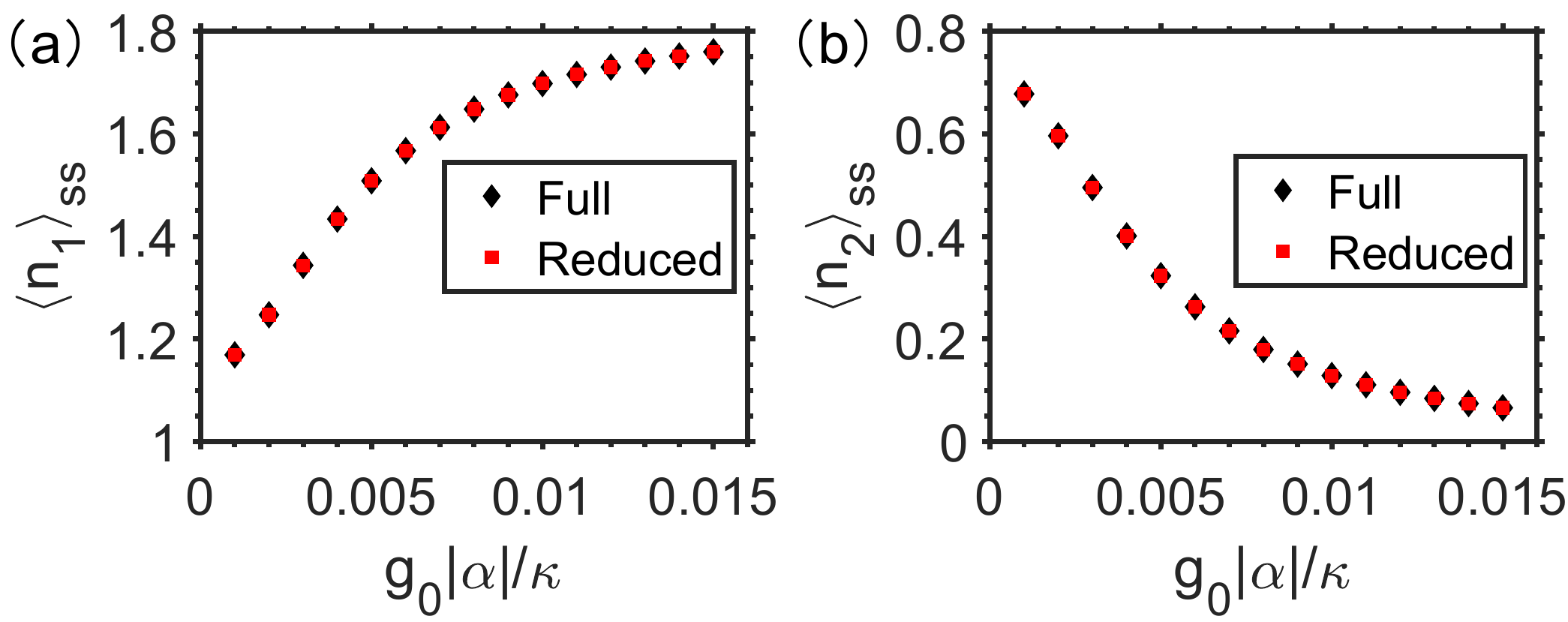}
    \caption{Comparison of numerical results obtained from the full master equation and the reduced master equation for the case of $N=2$. (a) The steady-state phonon number at the first mode. (b) The steady-state phonon number at the second mode. $T=1$ mK, $\omega_0=100$ MHz, $k/m=\omega_0^2/3$, $\kappa=1$ MHz, $\gamma_1=\gamma_2=100$ Hz, $\Delta=-k/(m\omega_0)$. The thermal phonon population is $\bar{n}_{1,\text{th}}=1.15, \bar{n}_{2,\text{th}}=0.71$. The cutoff of Fock state is set to 10 for phonons, and set to 5 for photon in the calculation of the master equations. }
    \label{fig:nvsg0acompared}
\end{figure}
\begin{widetext}
\begin{align}
   \dot{\rho}^I_m=&\tilde{\mathcal{L}}\rho^I_m\nonumber\\
   =&-i[\tilde{H},\rho^I_m]+\sum_iU_{i,i}^2\left\{\Gamma(2\omega_i)\mathcal{D}[\hat{b}_i\hat{b}_i]+\Gamma(-2\omega_i)\mathcal{D}[\hat{b}_i^{\dagger}\hat{b}_i^{\dagger}]\right\}\rho^I_m+\Gamma(0)\mathcal{D}\left[\sum_iU_{i,i}(\hat{b}_i^{\dagger}\hat{b}_i+\hat{b}_i\hat{b}_i^{\dagger})\right]\rho^I_m\nonumber\\
&+\sum_{i<j}4U_{i,j}^2\left\{\Gamma(\omega_i+\omega_j)\mathcal{D}[\hat{b}_i\hat{b}_j]+\Gamma(-\omega_i-\omega_j)\mathcal{D}[\hat{b}_i^{\dagger}\hat{b}_j^{\dagger}]+\Gamma(\omega_i-\omega_j)\mathcal{D}[\hat{b}_i\hat{b}_j^{\dagger}]+\Gamma(\omega_j-\omega_i)\mathcal{D}[\hat{b}_i^{\dagger}\hat{b}_j]\right\}\rho^I_m, \label{eq:reducedquantummasterequation}
\end{align}
\end{widetext}
where $\tilde{H}$ is the effective Hamiltonian induced by the interaction with the cavity field. The explicit form of $\tilde{H}$ is given by
\begin{widetext}
\begin{align}
    \tilde{H}=&\sum_{i=1}^NU_{i,i}^2\left[\tilde{\Delta}(2\omega_i)\hat{b}_i^{\dagger}\hat{b}_i^{\dagger}\hat{b}_i\hat{b}_i+\tilde{\Delta}(-2\omega_i)\hat{b}_i\hat{b}_i\hat{b}_i^{\dagger}\hat{b}_i^{\dagger}\right]+\tilde{\Delta}(0)\left[\sum_{i=1}^  NU_{i,i}(\hat{b}^{\dagger}_i\hat{b}_i+\hat{b}_i\hat{b}^{\dagger}_i)\right]^2\nonumber\\
    &+\sum_{i<j}4U_{i,j}^2\left[\tilde{\Delta}(\omega_i+\omega_j)\hat{b}^{\dagger}_i\hat{b}^{\dagger}_j\hat{b}_j\hat{b}_i+\tilde{\Delta}(-\omega_i-\omega_j)\hat{b}_i\hat{b}_j\hat{b}^{\dagger}_j\hat{b}^{\dagger}_i+\tilde{\Delta}(\omega_i-\omega_j)\hat{b}^{\dagger}_i\hat{b}_j\hat{b}^{\dagger}_j\hat{b}_i+\tilde{\Delta}(\omega_j-\omega_i)\hat{b}_i\hat{b}^{\dagger}_j\hat{b}_j\hat{b}^{\dagger}_i\right]
\end{align}
\end{widetext}
The energy shift $\tilde{\Delta}(\omega)$ and the rate $\Gamma(\omega)$ are given by
\begin{align}
    \tilde{\Delta}(\omega)=\text{Im}\left[G(\omega)\right],\quad \Gamma(\omega)=2\text{Re}\left[G(\omega)\right],
\end{align}
and $G(\omega)$ is the Fourier transform of the photon correlation function.
\begin{align}
    G(\omega)=\epsilon^2|\bar{\alpha}|^2\int_0^td\tau e^{i\omega\tau}e^{(i\Delta-\frac{\kappa}{2})\tau}.
\end{align}
Since $\tau$ integration is dominated by the times $\sim \kappa^{-1}$, which is much shorter than $t$, we can extend the $\tau$ integration to infinity,
\begin{align}
    G(\omega)=\epsilon^2|\bar{\alpha}|^2\int_0^{\infty}d\tau e^{i\omega\tau}e^{(i\Delta-\frac{\kappa}{2})\tau}=\frac{\epsilon^2|\bar{\alpha}|^2}{-i(\Delta+\omega)+\kappa/2}.
\end{align}
Note that $\rho_m^I(t)$ is still in the interaction picture. Transforming back to $\rho_m(t)$ we obtain
\begin{align}
    \dot{\rho}_m=\mathcal{L}_m\rho_m+\tilde{\mathcal{L}}\rho_m.
    \label{eq:appendixreduced}
\end{align}

The validity of the adiabatic elimination of the optical field in optomechanical systems have been verified in many numerical researches \cite{wilson2008cavity,lorch2014laser,wallquist2010single}. Here we give a simple comparison of the steady-state phonon number obtained from the full quantum master equation (Eq. (\ref{eq:fullmaster})) and the reduced quantum master equation (Eq. (\ref{eq:appendixreduced})) in the case of $N=2$, as shown in Fig. (\ref{fig:nvsg0acompared}). The good agreement shows the adiabatic approximation works very well in the (bad cavity, weak coupling) regime we are interested. 

From the master equation for the reduced density operator $\rho_m$, we can obtain the rate equations for the average phonon numbers, which is Eq. (\ref{eq:fullrateequation}) of the main text.
\begin{figure}[t]
    \centering
    \includegraphics[width=\linewidth]{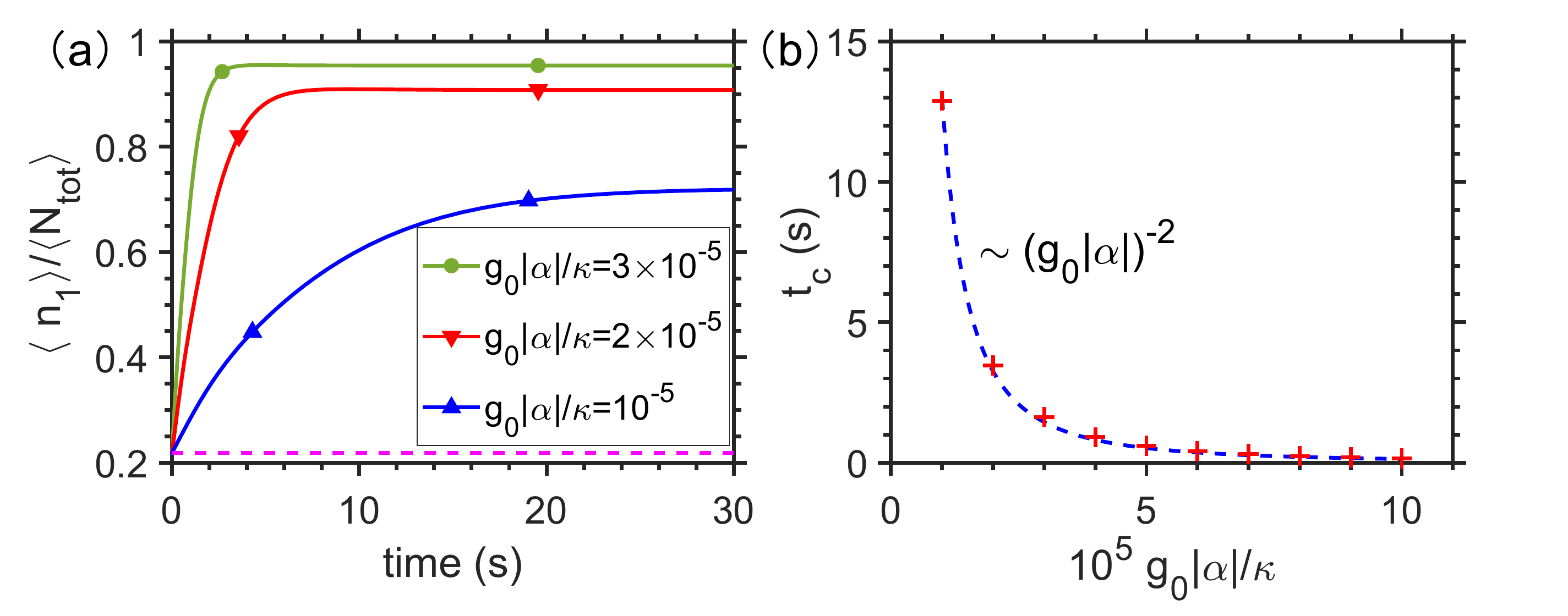}
    \caption{Time evolution of the ratio of the average phonon numbers at the lowest mode to the total phonon numbers. (a) Time evolution of $\langle \hat{n}_1\rangle/\langle \hat{N}_{\text{tot}}\rangle$ at different optomechanical couplings. (b) The time needed for $\langle \hat{n}_1\rangle/\langle \hat{N}_{\text{tot}}\rangle$ to achieve 90\% of the peak value with respect to optomechanical couplings. The dashed line in (a) denotes $\bar{n}_{1,\text{th}}/{N}_{\text{tot}}$ in thermal equilibrium. In (a) and (b), $N=5$, the cavity and membrane parameters are the same as those used in Fig. (\ref{fig:nvsdelta}).}
    \label{fig:nvst}
\end{figure}

\section{Comparison between Fr\"{o}hlich's original model and our model}
\label{appendix:comparison}
The rate equations of Fr\"{o}hlich's original model are
\begin{align}
    \dot{n}_l=&s-\phi(n_le^{\hbar\omega_l/KT}-(n_l+1))\nonumber\\
    &+\chi\sum_j[(n_l+1)n_j-n_l(1+n_j)e^{\hbar(\omega_l-\omega_j)/KT}] \label{eq:frohlichoriginal}
\end{align}
On the right-hand side of Eq. (\ref{eq:frohlichoriginal}), the first term describes the incoherent phonon pumping, the second term describes the dissipation induced by the heat bath, and the third term describes the two-phonon-energy-redistribution processes. The second term of Eq. (\ref{eq:frohlichoriginal}) is the same as the dissipation term of Eq. (\ref{eq:rateequation}) when we make use of the identity
\begin{align}
    e^{\hbar\omega_l/KT}=\frac{\bar{n}_{l,\text{th}}+1}{\bar{n}_{l,\text{th}}}
\end{align}

If there is no phonon pumping, i.e. $s=0$, the steady state of phonon distribution follows the thermal distribution. Thus, the main differences between Fr\"{o}hlich original model (Eq. (\ref{eq:frohlichoriginal})) and our model (Eq. (\ref{eq:rateequation})) are: (i) \emph{the incoherent phonon pumping is optional in our model, while the phonon pumping is necessary in Fr\"{o}hlich original model}; (ii) \emph{the two-phonon-energy-distribution processes in our model can be controlled by the driving laser, while these processes in Fr\"{o}hlich original model are determined by the environment}.

\section{Time evolution of average phonon numbers}
\label{appendix:timeevolution}
In Fig. (\ref{fig:nvst}a), we show the evolution of average phonon numbers at different optomechanical couplings in the case of $N=5$. It can be seen the evolution is faster for larger optomechanical couplings. The evolution rate of the average phonon numbers mainly depends on the energy-redistribution processes, i.e., the net transition rate $\Gamma_{j\to 1}$. In Fig. (\ref{fig:nvst}b), we show the time $t_c$ for $\langle \hat{n}_1\rangle/\langle \hat{N}_{\text{tot}}\rangle$ to reach 90\% of its peak value with respect to the optomechanical coupling. The numerical results fit well with the prediction $t_c\propto\Gamma_{j\to1}\propto g_0^2|\bar{\alpha}|^{-2}$.

\section{The case of multiple cavities}
\label{appendix:multiplecavities}
For the configuration where $M$ ($M\le N$) cavities are coupled to the first $M$ membranes, the Hamiltonian is
\begin{align}
    \hat{H}&=-\sum_{k=1}^M\Delta \hat{d}_k^{\dagger}\hat{d}_k+\sum_{j=1}^N\omega_j\hat{b}_j^{\dagger}\hat{b}_j+\frac{2g_0}{N+1}\sum_{k=1}^M(\bar{\alpha} \hat{d}_k^{\dagger}+\bar{\alpha}^*\hat{d}_k)\nonumber\\
    &\times\left[\sum_{j=1}^NU_{k,j,j}(\hat{b}_j+\hat{b}_j^{\dagger})^2+\sum_{i<j}2U_{k,i,j}(\hat{b}_i+\hat{b}_i^{\dagger})(\hat{b}_j+\hat{b}_j^{\dagger})\right], 
\end{align}
where
\begin{align}
    U_{k,i,j}=\frac{\omega_0\sin{\left(\frac{ki\pi}{N+1}\right)}\sin{\left(\frac{kj\pi}{N+1}\right)}}{\sqrt{\omega_i\omega_j}}. \label{eq:ukij}
\end{align}
For simplicity, we have assumed that the detuning, coupling and laser power are the same for all cavities. Simliar to the case of single cavity, the cavity modes can be eliminated via adiabatic approximation, 
\begin{widetext}
\begin{align}
   \dot{\rho}_m=&-i[\tilde{H}^{\prime},\rho_m]+\sum_{i}\left\{\gamma_i(1+\bar{n}_{i,\text{th}})\mathcal{D}[\hat{b}_i]+\gamma_i\bar{n}_{i,\text{th}}\mathcal{D}[\hat{b}_i^{\dagger}]\right\}\rho_m\nonumber\\
   &+\sum_{k=1}^{M}\sum_iU_{k,i,i}^2\left\{\Gamma(2\omega_i)\mathcal{D}[\hat{b}_i\hat{b}_i]+\Gamma(-2\omega_i)\mathcal{D}[\hat{b}_i^{\dagger}\hat{b}_i^{\dagger}]\right\}\rho_m+\sum_{k=1}^{M}\Gamma(0)\mathcal{D}\left[\sum_iU_{k,i,i}(\hat{b}_i^{\dagger}\hat{b}_i+\hat{b}_i\hat{b}_i^{\dagger})\right]\rho_m\nonumber\\
&+\sum_{k=1}^{M}\sum_{i<j}4U_{k,i,j}^2\left\{\Gamma(\omega_i+\omega_j)\mathcal{D}[\hat{b}_i\hat{b}_j]+\Gamma(-\omega_i-\omega_j)\mathcal{D}[\hat{b}_i^{\dagger}\hat{b}_j^{\dagger}]+\Gamma(\omega_i-\omega_j)\mathcal{D}[\hat{b}_i\hat{b}_j^{\dagger}]+\Gamma(\omega_j-\omega_i)\mathcal{D}[\hat{b}_i^{\dagger}\hat{b}_j]\right\}\rho_m, \label{eq:multiplereducedquantummasterequation}
\end{align}
\end{widetext}
Here the explicit form of the effective Hamiltonian $\tilde{H}^{\prime}$ is not given because it has no effect on the average phonon number. Hence, the rate equations for the average phonon numbers can be obtained
\begin{widetext}
\begin{align}
   \langle\dot{\hat{n}}_l\rangle=& -\gamma_l(\langle \hat{n}_l\rangle-\bar{n}_{l,\text{th}})+\sum_{k=1}^{M}\left[-2U_{k,l,l}^2\Gamma(2\omega_l)\langle \hat{n}_l(\hat{n}_l-1)\rangle+2U_{k,l,l}^2\Gamma(-2\omega_l)\langle(\hat{n}_l+1)(\hat{n}_l+2)\rangle\right]\nonumber\\
&+\sum_{k=1}^{M}\sum_{j\neq l}\left[-4U_{k,j,l}^2\Gamma(\omega_l+\omega_j)\langle \hat{n}_l\hat{n}_j\rangle+4U_{k,j,l}^2\Gamma(-\omega_l-\omega_j)\langle(\hat{n}_l+1)(\hat{n}_j+1)\rangle\right]\nonumber\\
&+\sum_{k=1}^{M}\sum_{j\neq l}\left[-4U_{k,j,l}^2\Gamma(\omega_l-\omega_j)\langle \hat{n}_l(\hat{n}_j+1)\rangle+4U_{k,j,l}^2\Gamma(\omega_j-\omega_l)\langle(\hat{n}_l+1)\hat{n}_j\rangle\right] . 
\end{align}
\end{widetext}
 By introducing the coefficients
\begin{align}
    \tilde{U}_{i,j}^2=\sum_{k=1}^MU_{k,i,j}^2
    \label{eq:effectiveu}
\end{align}
we get rate equations for average phonon numbers in the same form as Eq. (\ref{eq:fullrateequation}) except that $U_{i,j}^2$ is replaced by $\tilde{U}_{i,j}^2$. We consider the case $M=N$, i.e., every membrane is coupled to a corresponding cavity. Combining Eq. (\ref{eq:ukij}) and Eq. (\ref{eq:effectiveu}), we can simplify the coefficients as
\begin{align}
    \tilde{U}_{i,j}^2=\frac{(N+1)\omega_0^2}{\omega_i\omega_j}\times\begin{cases}
    \frac{3}{8}, & \text{for $i=j$}.\\
    \frac{1}{4}, & \text{for $i\neq j$}.
  \end{cases}
\end{align}
which is Eq. (\ref{eq:effectiveuij}) in the main text.


%

\end{document}